# Deep Learning Based Image Retrieval in the JPEG Compressed Domain

Shrikant Temburwar[1][0000-0001-5831-3825], Bulla Rajesh[2][0000-0002-5731-9755]*
and Mohammed Javed[3][0000-0002-3019-7401]

Computer Vision and Biometrics Laboratory (CVBL)
Department of Information Technology
Indian Institute of Information Technology Allahabad, Prayagraj, U.P, India
[1]shrikant.temburwar@gmail.com, [2]rajesh091106@gmail.com, [3]javed@iiita.ac.in

**Abstract.** Content-based image retrieval (CBIR) systems on pixel domain use low-level features, such as colour, texture and shape, to retrieve images. In this context, two types of image representations i.e. local and global image features have been studied in the literature. Extracting these features from pixel images and comparing them with images from the database is very time-consuming. Therefore, in recent years, there has been some effort to accomplish image analysis directly in the compressed domain with lesser computations. Furthermore, most of the images in our daily transactions are stored in the JPEG compressed format. Therefore, it would be ideal if we could retrieve features directly from the partially decoded or compressed data and use them for retrieval. Here, we propose a unified model for image retrieval which takes DCT coefficients as input and efficiently extracts global and local features directly in the JPEG compressed domain for accurate image retrieval. The experimental findings indicate that our proposed model performed similarly to the current DELG model which takes RGB features as an input with reference to mean average precision while having a faster training and retrieval speed.

**Keywords:** Image Retrieval, Compressed Domain, DCT Coefficients, Unified Model.

## 1 Introduction

"Content-Based Image Retrieval" (CBIR) aims to retrieve images that match precise criteria for colour, shape and texture of a given query image. In the CBIR system, the input is a selection of seed images or colour/texture scales as examples, and the retrieval system attempts to match images in the database with test patterns based on the above visual characteristics. Generally, images are saved in a compressed state and need to be decompressed for feature extraction. Conventionally, image retrieval on JPEG images requires first decompressing the image and then searching in the spatial domain. This makes the decompression process very time consuming, particularly for huge image databases, and therefore computationally and processing time intensive. With the development of compression standards, images in JPEG format alone account for more than 95% of the images on the Internet [1]. For this reason, CBIRs implemented directly in the "JPEG compressed domain" has garnered a great deal of interest. Retrieval in the compressed domain is an attempt to extract feature vectors directly from compressed or partially decoded data. This can significantly improve the processing efficiency while reducing the computer resource requirements.

*Corresponding Author

To achieve high-performance image retrieval, two types of image representations are required: "global and local features". Global features [2,8,5,7,19,20], also called "global descriptors" or "embeddings", summarise the content of an image and usually result in a covenant representation. However, details on visual element's spatial structure are lost. In contrast, local features [10,9,21,6,14,2] consist of descriptors of specific image regions and geometric information and are particularly useful for matching images that describe rigid bodies. In general, "global features" are good at reproducing and "local features" are good at accuracy. Local features cannot learn the resemblance of very different poses, whereas global features do. However, geometric validation based on local features generally yields scores that reflect the similarity of images well and are more reliable than the distances of global features.

A retrieval system's general strategy is to fetch global features first, then use local feature matching to identify the best images in the database [2,6]. Many systems that depend on both features currently have to extract each feature independently using different models. If these two models require specialised and restricted hardware, such as GPUs, this is undesirable because it increases memory consumption and latency. Furthermore, equivalent measurements are often conducted on both models, resulting in redundant processing and excessive complexity. Therefore, as reported in the literature [2], a unified model for extracting both global and local features will be used in this present work.

In this paper, we propose a deep learning-based system for image retrieval in the compressed domain that uses DCT coefficients as input and needs only slight changes to the current DELG [2] model that uses RGB input. We validated our method on the ROxf dataset [18] and managed to train faster than the baseline model. To the best of our understanding, this is the first study to look at the role of image retrieval in a compressed domain using Deep Neural Networks. The experimental findings indicate that our proposed model performed similarly to the current DELG model with reference to mean average precision while having a faster training and retrieval speed. The remainder of this paper is arranged as follows- section 2 covers relevant work in the field of CBIR, section 3 explains the background methodology and proposed model, section 4 gives experimental methods and results, and section 5 summarises the paper briefly.

## 2  Related Work

### 2.1  Deep Local and Global features (DELG)

The usage of two different models, one for image representation and one for local descriptor extraction, is incompatible with the finite resources and performance standards available in many applications [33]. As a result, some researchers are looking at the hybrid model that incorporates the computation of local descriptors for spatial validation and global descriptors for similarity comparison in a multi-headed CNN. This technique is used in DELG [2], which extracts all global and local features from a similar backbone containing two heads: i) "GeM pooling" [19], which generates global representations, and ii) "Attention module local descriptors", which is inspired by DELF [6]. The authors employed a hierarchical representation of CNNs to train the two tasks synchronously [17]. Local features are bound to the intermediate layers and encode more local content, while global features are bound to the deeper network layers and encode high-level cues. As a result, only the similarity loss gradients of the global descriptors are sent to the backbone during training, while the loss gradients associated with the local descriptors are terminated early. This is because naive triple-loss optimisation distorts the hierarchical representation of features and yields weak models.

### 2.2    Deep Local Features (DELF)

DELF [6] employs coarse-region features of a pre-trained CNN's convolutional layer to train a smaller CNN for measuring the significance of the closely sampled main points. Prior to training, the weights of these projections are used to weight the local descriptors and pool them into a global feature vector, enabling image-level tracking to fine-tune local features.

### 2.3    Discrete Cosine Transform in Computer Vision

In classical computer vision algorithms, the "Discrete Cosine Transform" is commonly used [24,25,26,22,30,15,35] to encode RGB images in the "spatial domain" into the "frequency domain" components. Several studies have been conducted to incorporate DCT into deep learning-based computer vision frameworks: "Ulicny et al [27] used CNNs to interpret DCT encoded images. Ehrlich et al [29] suggested ResNet in the DCT domain". The semantine segmentation of DCT representation by Lo et al. [28] was re-ordered and passed on the DCT coefficients for a CNN; "Xu et al investigated learning in the frequency domain for target recognition and instance segmentation, using DCT coefficients as input for their models instead of RGB input for the CNN model" [3]. They proposed a frequency domain learning method that uses frequency domain information as input and has the same popular structure such as "ResNet-50", "MobileNetV2", and "Mask R-CNN". Their experimental results show that learning with static channel selection in the frequency domain could achieve greater accuracy than traditional spatial downsampling methods, while also greatly reducing the size of input data. Specifically, with the same input scale ImageNet classification, their proposed approach improved top 1 accuracy by 1.60% and 0.63 % on "ResNet-50" and "MobileNetV2", respectively. Also, their proposed approach boosts top 1 accuracy by 1.42 % with half the input size.

This research paper uses a similar approach as discussed above for image retrieval in the JPEG compressed domain.

## 3    Methodology

### 3.1    The JPEG Encoder

"JPEG is the most widely used image compression technique and accepted as the ISO standard for still image coding. It is built on the Discrete Cosine Transform (DCT), a variant of the Discrete Fourier Transform" [23,31,34]. As shown in Figure 1 [15], JPEG encoding consists of the following stages. "First, an (RGB) image is usually converted into the YCbCr space. The reason for this is that the human visual system is less sensitive to changes in the chrominance (Cb and Cr) than in the luminance (Y) channel. Consequently, the chrominance channels can be downsampled by a factor of 2 without sacrificing too much image quality, resulting in a full resolution Y and downsampled Cb and Cr components.

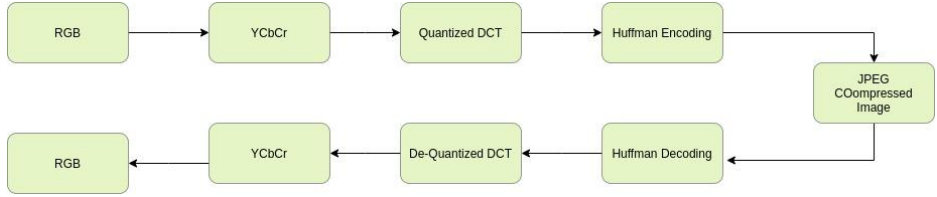

**Fig. 1.** JPEG compression and decompression flow diagram [15].

The image is then divided (each colour channel separately) into 8 × 8 pixel sub-blocks and DCT applied to each such block. The 2-d DCT for an 8 × 8 block $f_{x,y}$, x, y = 0 . . . 7 is defined as in equation 1.

$$F_{u,v} = \frac{C_u C_v}{4} \sum_{x=0}^{7} \sum_{y=0}^{7} f_{x,y} \cos\left[\frac{\pi(2x+1)u}{16}\right] \cos\left[\frac{\pi(2y+1)v}{16}\right] \quad (1)$$

where $C_u, C_v = 1/\sqrt{2}$ for $u, v = 0$, otherwise $C_u, C_v = 1$. $u$ and $v$ are horizontal and vertical spatial frequency respectively. $f_{x,y}$ is a value of pixel at (x, y) and $F_{u,v}$ is DCT coefficient at (u, v).

The majority of the information in all DCT coefficients is contained in a few low-frequency coefficients. Out of 64 DCT coefficients, zero-frequency coefficient is referred to as a DC coefficient while the other 63 are referred to as AC coefficients. The DC term represent the mean of the picture block while the higher frequencies are determined by the AC coefficients. Higher frequencies can be ignored since lower frequencies are more appropriate for image content. JPEG accomplishes this by using a quantisation step that loosely quantises higher frequencies while more precisely retaining lower frequencies" [34].

Because the DC terms fluctuate slowly throughout the picture, they are differently coded after quantization. The AC coefficients are runlength coded and are arranged in a zig-zag pattern for every block. Finally, to increase compression efficiency, both portions are entropy (Huffman) coded. Decoding or decompressing the JPEG requires a subsequent inverse transformation, which is done in the opposite order of the previous steps. All these steps are lossless excluding quantization inverse transformation. The restored image is blurred to some degree due to the lack of precision in the DCT coefficient quantization operation.

### 3.2 Proposed Model

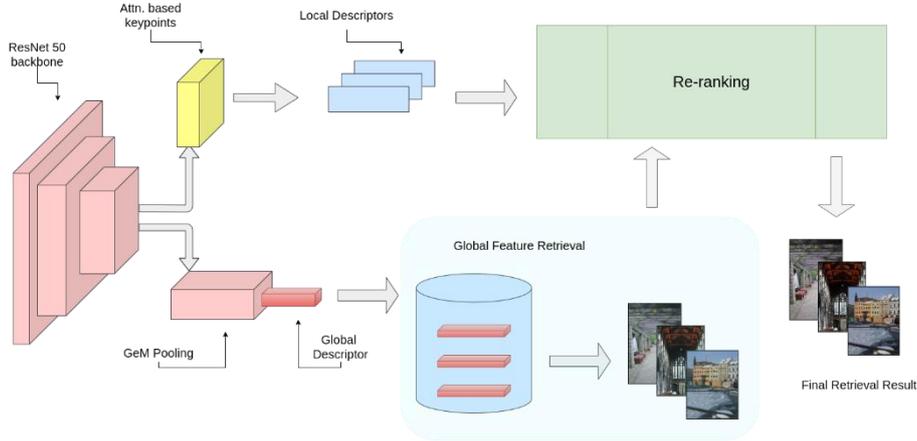

**Fig. 2.** The block diagram of the proposed model [2].

Figure 2 depicts our proposed model, which is identical to DELG [2] ("DEep Local and Global features") except for the removal of the autoencoder module. This model (on the left) extracts both "local and global deep features". To quickly find the most relevant images, global features can be used in the first phase of the search method (bottom). To increase the system's accuracy, local features can be used to adjust the ranking of the best outcomes (top right). The unified model learns global and local features using a layered representation of convolutional neural networks and integrates recent developments such as global pooling and attuned local feature recognition. For reflecting the various features types to be studied, the proposed model employs hierarchical representations from CNNs. Although global features are appropriate for deep layers that reflect cues at the highest level, local features are better suited for middle layers that gets the localised content.

We use "Generalised mean pooling" (GeM) [19] to combine "Deep activations" into a global feature, which essentially weights each feature's contributions. Another important aspect of learning a global feature is whitening the aggregated representation, which we incorporate into the model through a "fully-connected layer F" with bias b to produce global feature. When it comes to "local features", it is critical to pick only the appropriate matching regions. That can be accomplished by employing an "Attention module M" [6], the purpose of which is to predict discriminated "local features" which are derived for points of interest.

## 4 Experiments and Results

### 4.1 Pre-processing of Data in the Compressed Domain

Figure 3 depicts the flow of pre-processed data, in which we adopt a pre-processing and enlargement flow in the spatial domain [3], which includes image resizing, cropping, and flipping. The image is then transformed to the DCT domain after being transferred to the YCbCr colour space. A 3D DCT cube is formed in the flow by grouping 2D DCT coefficients with the same frequency. Following that, a subset of prominent DCT channels is chosen. A tensor was generated by concatenating the chosen channels in the YCbCr colour space. Finally, each DCT channel is normalised using the mean and variance determined on training data.

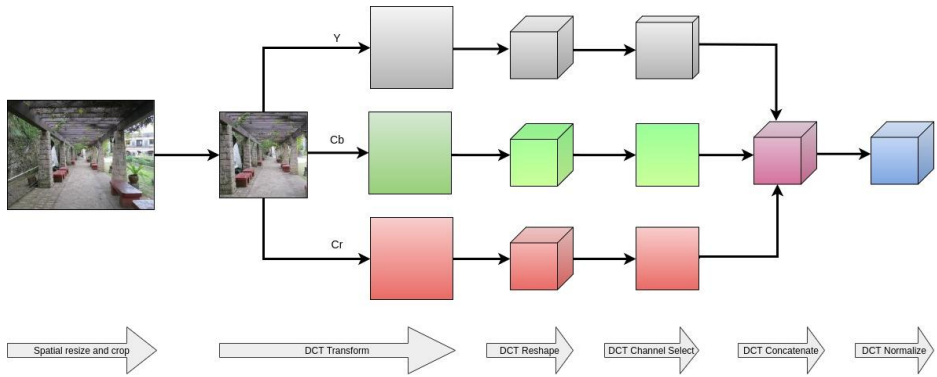

**Fig. 3.** The pipeline for pre-processing data [3].

Figure 3 depicts the flow of pre-processed data, in which we adopt a pre-processing and enlargement flow in the spatial domain [3], which includes image resizing, cropping, and flipping. The image is then transformed to the DCT domain after being transferred to the YCbCr colour space. A 3D DCT cube is formed in the flow by grouping 2D DCT coefficients with the same frequency. Following that, a subset of prominent DCT channels is chosen. A tensor was generated by concatenating the chosen channels in the YCbCr colour space. Finally, each DCT channel is normalised using the mean and variance determined on training data.

The 2D DCT coefficients are then reshaped into a 3D DCT cube by the DCT reshaping process. "We grouped components of the same frequency into all $8 \times 8$ blocks in a channel, maintaining their spatial relationship at each frequency, since the JPEG compression standard uses an $8 \times 8$ DCT transform on the YCbCr colour space. As a result, each Y, Cb, and Cr components provides $8 \times 8 = 64$ channels, one at each frequency, which gives a total of 192 channels. The original RGB input image is assumed to have the form $H \times W \times C$, where $C = 3$, and the image's height and width are denoted by H and W. The shape of the input features changes to $H / 8 \times W / 8 \times 64C$ after conversion to the frequency domain, so the size of the input data remains constant" [3,4].

We bypass the stride-2 convolution input layer of the standard CNN model since the input function map in the DCT domain is smaller than its equivalent in the spatial domain. We also skip the maximum pooling operator if it follows the input convolution immediately. The next layer channel size is then fine-tuned based on the number of channels in the DCT domain. Figure 4 depicts this [3]. The three input layers of "ResNet-50" [13] are eliminated to make a "$56 \times 56 \times 64$" DCT input. We change the original CNN model in this way to accept DCT features as input.

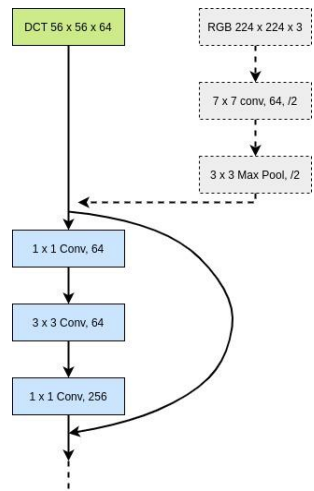

**Fig. 4.** Attaching the pre-processed DCT features input to ResNet-50 [3].

CNN models commonly use $224 \times 224 \times 3$ input data in image classification tasks, which is typically downsampled from far higher resolution images. "In ResNet50, for example, the input DCT features are bound to the first residual block, raising the number of channels to 192, resulting in an input feature of the form $56 \times 56 \times 192$, as seen in Figure 4, which is a DCT transform of the input image with a size of $448 \times 448 \times 3$, retaining four times more information than the corresponding input feature of $224 \times 224 \times 3$ in the spatial domain" [3].

### 4.2 Static Frequency Channel Selection

The low-frequency and luminance components of the JPEG compression standard are given more bits [3]. The low-frequency channels are statically selected using the same convention to emphasise the Lumi component over the colour component. The frequency channels with the highest activation probability are then introduced to the CNN model. The image codec will trim the remaining channels to reduce the amount of bandwidth available for data transmission and the size of the input data.

### 4.3 Model Implementation

The model has been implemented in PyTorch. As a backbone, we used ResNet-50 (R50) [13]. The GeM pooling [19] with parameter p = 3 and a two-convolutional layer attention network with no stride and a kernel size of one is used. We use ReLU in first layer and Softplus [11] in second layer as activation functions.

### 4.4 Training Details

We used a small subset of the Google Landmarks dataset v2 (GLDv2) [6,36] with 17 classes, which we divided into two training : validation subsets with an 80 : 20 split. The training split is used to train the classifier, while the validation split is used to validate the trained model as training continues. The model is initialized using pre-trained ImageNet weights. The image is enlarged arbitrarily by changing the aspect ratio, and then reduced to a resolution of 448 × 448. We trained the model for 40 epochs using the batch size of 32 on a single "Tesla T4 GPU" with 2 core CPU on Google Colab. "The Stochastic Gradient Descent (SGD) optimizer was employed. We initialized the model with a learning rate of 0.1 and with a momentum of 0.9. We also initialize weight decay of 1e-4 for an interval of 10 epochs. The ArcFace margin is set to m = 0.15, and the attention loss weights are set to $\beta$ = 1. The global loss learnable scalar is initialised to $\gamma$ = 30" [2].

### 4.5 Evaluation Dataset

To evaluate, we used the Oxford dataset [17] with modified annotations, which we call ROxf [18]. The ROxf dataset comprises 4993 database images, each with a unique query set, and 70 images. The "mean average precision (mAP)" is used for performance evaluation.

"The labels are (easy, hard, unclear) positively or negatively treated or ignored by three evaluation settings with different degrees of difficulty.
- **Easy (E):** Easy images are taken as positive and difficult or unclear images are ignored.
- **Medium (M):** Images that are easy or difficult are viewed positively, and unclear images are ignored.
- **Hard (H):** Hard images are viewed positively, whereas images that are easy and unclear are ignored" [18].

If no positive images are found for a query of a specific configuration, the query is excluded from evaluation.

### 4.6 Feature Extraction and Matching

Following the same conventions as DELG and DELF [2,6,19,7], we use image pyramids with inference time to generate a multi-scale representation. We used a single scale for global features and experimented with a scale of $\{1 / \sqrt{2}, 1, \sqrt{2}\}$ for local features. "The attention score A is used to select local features. A limit of 200 local features with Attention score $\tau$ are permitted, where $\tau$ is set to the median of the attention scores in the previous training iteration. To fit local features, we use RANSAC [12] with an affine model. The top 100 ranked images in the first phase are taken into consideration while re-ranking the global feature search results based on local features" [2].

### 4.7 Results

The RGB image of resolution of 448 × 448 with 3 channels is processed to get DCT features of size 56 × 56 with 64 channels. These DCT features are subsequently fed into our training model as an input. The trained model is then used to extract both "global and local features" from the testing dataset for image retrieval. We also trained DELG [2] model which takes RGB as an input on the same data and performed retrieval on it for comparison. The images are retrieved first by global features and then ranked by matching local features and spatial verification. The results of the image retrieval on the ROxf dataset are shown in Table 1 and 2. The values depict the percentage mean average precision of retrieved images for given Easy, Medium, Hard queries respectively.

**Table 1.** Comparison between RGB and DCT based model.

| Features | Channels | Size / Channel | Training Time (Per Epoch) |
|---|---|---|---|
| RGB [2] | 3 | $224 \times 224$ | 196 Sec |
| DCT | 64 | $56 \times 56$ | 162 Sec |

**Table 2.** Results for the retrieval tasks on the ROxf dataset

| Features | mAP E (%) | mAP M (%) | mAP H (%) |
|---|---|---|---|
| RGB Global [2] | 18.46 | 16.14 | 5.96 |
| RGB Global + Local [2] | 32.69 | 23.89 | 7.44 |
| DCT Global | 20.34 | 17.59 | 6.09 |
| DCT Global + Local | 30.16 | 25.42 | 11.3 |

It is observed that even passing $1.33 \times$ larger DCT features as an input boost the training speed by $1.2 \times$ compared to passing a smaller RGB features as input. It is also been observed that ranking local features increases our model's performance substantially compared to retrieval using only global features. Figure 5 shows five most similar retrieved images from query dataset.

Due to insufficient hardware resources, we trained our model on a small dataset so the results we got seems much lower compared to the results on existing DELG and other retrieval models. Our model may perform as per DELG benchmark if trained on similar bigger dataset.

## 5 Conclusion

We propose a method for image retrieval in the compressed domain that is versatile and superior for image retrieval tasks. Since it needs few modifications to existing DELG models that use RGB inputs, our methodology can be extended to existing network training and inference methods. In the compressed domain, we obtained a faster training and retrieval speed than the baseline model, which takes RGB inputs. This has the potential to greatly increase computing performance while also lowering the computational resource needs for training larger datasets. As future work, we intend to expand the validation of the proposed model to the larger database such GLDv2 and compare the results by using different model backbones such as "ResNet-101, ResNet-110, ResNet-152" and other neural network architectures.

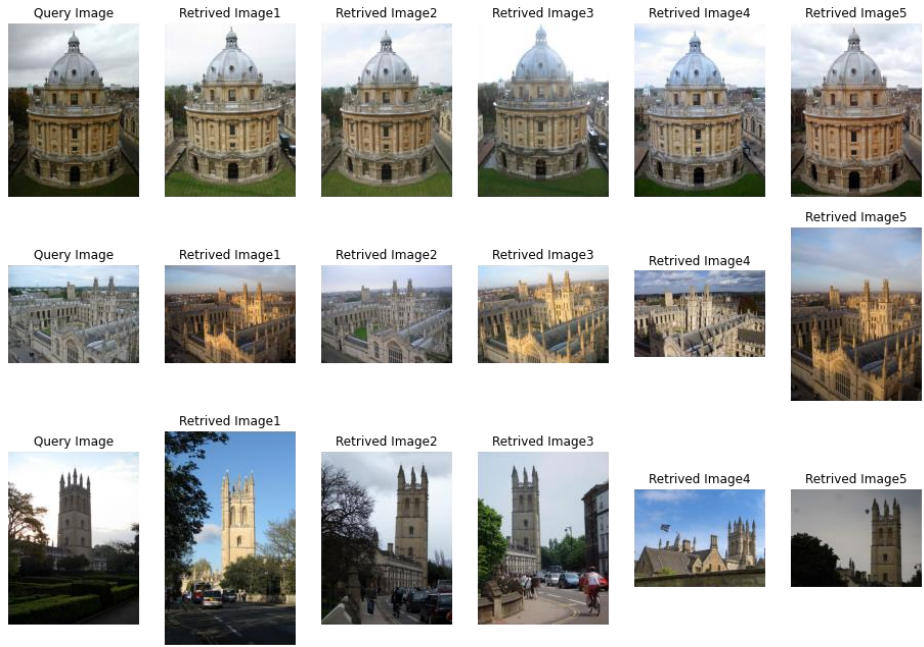

**Fig. 5.** Five most similar images for a given query image (shown row wise)
**References**

1. Jiang, J., Armstrong, A. and Feng, G.C.: Direct content access and extraction from JPEG compressed images. Pattern Recognition, 35(11), pp. 2511-2519 (2002).
2. Cao, B., Araujo, A. and Sim, J.: Unifying deep local and global features for image search. In European Conference on Computer Vision, pp. 726-743. Springer, Cham (2020).
3. Xu, K., Qin, M., Sun, F., Wang, Y., Chen, Y.K. and Ren, F.: Learning in the frequency domain. In Proceedings of the IEEE/CVF Conference on Computer Vision and Pattern Recognition, pp. 1740-1749 (2020).
4. Gueguen, L., Sergeev, A., Kadlec, B., Liu, R. and Yosinski, J.: Faster neural networks straight from jpeg. Advances in Neural Information Processing Systems, 31, pp. 3933-3944 (2018).
5. Arandjelovic, R., Gronat, P., Torii, A., Pajdla, T. and Sivic, J.: NetVLAD: CNN architecture for weakly supervised place recognition. In Proceedings of the IEEE conference on computer vision and pattern recognition, pp. 5297-5307 (2016).
6. Noh, H., Araujo, A., Sim, J., Weyand, T. and Han, B.: Large-scale image retrieval with attentive deep local features. In Proceedings of the IEEE international conference on computer vision, pp. 3456-3465 (2017).
7. Gordo, A., Almazan, J., Revaud, J. and Larlus, D.: End-to-end learning of deep visual representations for image retrieval. International Journal of Computer Vision, 124(2), pp. 237-254 (2017).

8. Jégou, H., Perronnin, F., Douze, M., Sánchez, J., Pérez, P. and Schmid, C.: Aggregating local image descriptors into compact codes. IEEE transactions on pattern analysis and machine intelligence, 34(9), pp. 1704-1716 (2011).
9. Bay, H., Ess, A., Tuytelaars, T. and Van Gool, L.: Speeded-up robust features (SURF). Computer vision and image understanding, 110(3), pp. 346-359 (2008).
10. Lowe, D.G.: Distinctive image features from scale-invariant keypoints. International journal of computer vision, 60(2), pp. 91-110 (2004).
11. Dugas, C., Bengio, Y., Bélisle, F., Nadeau, C. and Garcia, R.: Incorporating second-order functional knowledge for better option pricing. Advances in neural information processing systems, pp. 472-478 (2001).
12. Fischler, M.A. and Bolles, R.C.: Random sample consensus: a paradigm for model fitting with applications to image analysis and automated cartography. Communications of the ACM, 24(6), pp. 381-395 (1981).
13. He, K., Zhang, X., Ren, S. and Sun, J.: Deep residual learning for image recognition. In Proceedings of the IEEE conference on computer vision and pattern recognition, pp. 770-778 (2016).
14. Mishkin, D., Radenovic, F. and Matas, J.: Repeatability is not enough: Learning affine regions via discriminability. In Proceedings of the European Conference on Computer Vision (ECCV), pp. 284-300 (2018).
15. Rajesh, B., Javed, M. and Srivastava, S.: Dct-compcnn: A novel image classification network using jpeg compressed dct coefficients. In 2019 IEEE Conference on Information and Communication Technology, pp. 1-6. IEEE (2019).
16. Ozaki, K. and Yokoo, S.: Large-scale landmark retrieval/recognition under a noisy and diverse dataset. arXiv preprint arXiv:1906.04087 (2019).
17. Philbin, J., Chum, O., Isard, M., Sivic, J. and Zisserman, A.: Object retrieval with large vocabularies and fast spatial matching. In 2007 IEEE conference on computer vision and pattern recognition, pp. 1-8. IEEE (2007).
18. Radenović, F., Iscen, A., Tolias, G., Avrithis, Y. and Chum, O.: Revisiting oxford and paris: Large-scale image retrieval benchmarking. In Proceedings of the IEEE Conference on Computer Vision and Pattern Recognition, pp. 5706-5715 (2018).
19. Radenović, F., Tolias, G. and Chum, O.: Fine-tuning CNN image retrieval with no human annotation. IEEE transactions on pattern analysis and machine intelligence, 41(7), pp. 1655-1668 (2018).
20. Revaud, J., Almazán, J., Rezende, R.S. and Souza, C.R.D.: Learning with average precision: Training image retrieval with a listwise loss. In Proceedings of the IEEE/CVF International Conference on Computer Vision, pp. 5107-5116 (2019).
21. Yi, K.M., Trulls, E., Lepetit, V. and Fua, P.: Lift: Learned invariant feature transform. In European conference on computer vision, pp. 467-483. Springer, Cham (2016).
22. Javed, M., Nagabhushan, P. and Chaudhuri, B.B.: A review on document image analysis techniques directly in the compressed domain. Artificial Intelligence Review, 50(4), pp. 539-568 (2018).
23. Hudson, G., Léger, A., Niss, B. and Sebestyén, I.: JPEG at 25: Still going strong. IEEE MultiMedia, 24(2), pp. 96-103 (2017).
24. Shen, X., Yang, J., Wei, C., Deng, B., Huang, J., Hua, X.S., Cheng, X. and Liang, K.: Dct-mask: Discrete cosine transform mask representation for instance segmentation. In Proceedings of the IEEE/CVF Conference on Computer Vision and Pattern Recognition, pp. 8720-8729 (2021).
25. Chadha, A.R., Vaidya, P.P. and Roja, M.M.: Face recognition using discrete cosine transform for global and local features. In 2011 International Conference On Recent Advancements In Electrical, Electronics And Control Engineering, pp. 502-505. IEEE (2011).

26. Ravì, D., Bober, M., Farinella, G.M., Guarnera, M. and Battiato, S.: Semantic segmentation of images exploiting DCT based features and random forest. Pattern Recognition, 52, pp. 260-273 (2016).
27. Ulicny, M. and Dahyot, R.: On using cnn with dct based image data. In Proceedings of the 19th Irish Machine Vision and Image Processing conference IMVIP (Vol. 2) (2017).
28. Lo, S.Y. and Hang, H.M.: Exploring semantic segmentation on the DCT representation. In Proceedings of the ACM Multimedia Asia, pp. 1-6 (2019).
29. Ehrlich, M. and Davis, L.S.: Deep residual learning in the jpeg transform domain. In Proceedings of the IEEE/CVF International Conference on Computer Vision, pp. 3484-3493 (2019).
30. Rajesh, B., Javed, M. and Nagabhushan, P.: Automatic tracing and extraction of text-line and word segments directly in JPEG compressed document images. IET Image Processing, 14(9), pp. 1909-1919 (2020).
31. Murshed, M., Teng, S.W. and Lu, G.: Cuboid segmentation for effective image retrieval. In 2017 International Conference on Digital Image Computing: Techniques and Applications (DICTA), pp. 1-8. IEEE (2017).
32. Suresh, P., Sundaram, R.M.D. and Arumugam, A.: Feature extraction in compressed domain for content based image retrieval. In 2008 International Conference on Advanced Computer Theory and Engineering, pp. 190-194. IEEE (2008).
33. Masone, C. and Caputo, B.: A survey on deep visual place recognition. IEEE Access, 9, pp. 19516-19547 (2021).
34. Schaefer, G., Edmundson, D. and Sakurai, Y.: Fast JPEG image retrieval based on AC Huffman tables. In 2013 International Conference on Signal-Image Technology & Internet-Based Systems, pp. 26-30. IEEE (2013).
35. Rajesh, B., Javed, M., Nagabhushan, P. and Osamu, W.: Segmentation of text-lines and words from JPEG compressed printed text documents using DCT coefficients. In 2020 Data Compression Conference (DCC), pp. 389-389. IEEE (2020).
36. Weyand, T., Araujo, A., Cao, B. and Sim, J.: Google landmarks dataset v2-a large-scale benchmark for instance-level recognition and retrieval. In Proceedings of the IEEE/CVF Conference on Computer Vision and Pattern Recognition, pp. 2575-2584 (2020).